\documentclass[sigconf,screen]{acmart}

\AtBeginDocument{%
  \providecommand\BibTeX{{%
    \normalfont B\kern-0.5em{\scshape i\kern-0.25em b}\kern-0.8em\TeX}}}

\setcopyright{acmlicensed}
\acmPrice{15.00}
\acmDOI{10.1145/3540250.3560877}
\acmYear{2022}
\copyrightyear{2022}
\acmSubmissionID{fse22ivr-p1-p}
\acmISBN{978-1-4503-9413-0/22/11}
\acmConference[ESEC/FSE '22]{Proceedings of the 30th ACM Joint European Software Engineering Conference and Symposium on the Foundations of Software Engineering}{November 14--18, 2022}{Singapore, Singapore}
\acmBooktitle{Proceedings of the 30th ACM Joint European Software Engineering Conference and Symposium on the Foundations of Software Engineering (ESEC/FSE '22), November 14--18, 2022, Singapore, Singapore}

\usepackage{tabularx}

\begin{document}

\title{Paving the Way for Mature Secondary Research: The Seven Types of Literature Review}

\author{Paul Ralph}
\affiliation{%
  \institution{Dalhousie University}
  \city{Halifax}
  \country{Canada}}
\email{paulralph@dal.ca}

\author{Sebastian Baltes}
\affiliation{%
  \institution{University of Adelaide}
  \city{Adelaide}
  \country{Australia}}
\email{sebastian.baltes@adelaide.edu.au}

\renewcommand{\shortauthors}{Ralph and Baltes}

%
\begin{abstract}
  Confusion over different kinds of secondary research, and their divergent purposes, is undermining the effectiveness and usefulness of secondary studies in software engineering. This short paper therefore explains the differences between \textit{ad hoc review, case survey, critical review, meta-analysis} (aka systematic literature review), \textit{meta-synthesis} (aka thematic analysis), \textit{rapid review} and \textit{scoping review} (aka systematic mapping study). These definitions and associated guidelines help researchers better select and describe their literature reviews, while helping reviewers select more appropriate evaluation criteria.
\end{abstract}



\maketitle

\section{Introduction}

Scholarship and scholarly writing can be stratified into \textit{primary}, \textit{secondary}, and \textit{tertiary} research. 

\textit{Primary} research involves making observations in the broadest sense of collecting data about objects that are not themselves studies. Computing code metrics, administering questionnaires, interviewing participants, counting bugs, collecting documents, downloading source code, and taking field notes while observing a retrospective meeting are all primary research.

\textit{Secondary} research involves analyzing, synthesizing and critiquing primary studies. Ad hoc reviews, case surveys, critical reviews, meta-analysis, meta-synthesis and scoping reviews are all types of secondary research. 
Secondary research is central to evidence-based practice~\cite{kitchenham2004evidence} because (1) practitioners cannot read every study; (2) important decisions typically should be made based on the balance of evidence rather than a single study; (3) in many fields, individual studies are too small to produce accurate estimates of population parameters (e.g. the strength of the relationship between two variables). However, some secondary research degrades into predominately descriptive ``papers about papers'' with limited scope and usefulness.

\textit{Tertiary} scholarship has two related meanings: (1) analyses of groups of secondary studies (meta-reviews); (2) summaries or indices of broad areas of scientific knowledge as found in textbooks, encyclopedia entries, etc. 

We focus on secondary research because confusion over different kinds of secondary research, and their divergent purposes, undermines the effectiveness and usefulness of secondary studies in software engineering (SE), which motivates the following objective:

\smallskip
{\narrower \noindent \textit{\textbf{Objective:} The purpose of this article is to compile, describe, compare, and contrast the different kinds of secondary research commonly published in software engineering venues, to pave the way for more mature secondary research.} \par}
\smallskip

\begin{table*}
  \caption{The Seven Types of Literature Review}
  \label{tab:types}
  \footnotesize
\begin{tabular*}{\linewidth}{ l l l l l l}
\toprule
Type & Systematic & Purpose & Primary Studies & Analysis & Approach \\
\midrule
ad hoc & no & discuss & any & any & discuss purposively-selected related work \\ 
case \mbox{survey} & yes & explain \& predict & qualitative & quantitative & test causal hypotheses by aggregating case study results \\
critical & yes & prescribe & any & any & defend a position and make recommendations by analyzing a sample of papers \\
meta-analysis & yes & explain \& predict & quantitative & quantitative & estimate effect sizes by aggregating results of similar quantitative studies \\
meta-synthesis & yes & explain & qualitative & qualitative & synthesize the findings of numerous studies using qualitative analysis \\
rapid & yes & explain \& predict & quantitative & quantitative & a meta-analytic review that compromises rigor for speed \\
scoping & yes & describe & any & both & describe an area of research and map studies into meaningful categories \\
\bottomrule
\end{tabular*}
\end{table*}

\vspace{-0.5ex}

\section{Review Types}

\subsection{Ad Hoc Reviews}

An \textit{ad hoc} literature review is simply a discussion of some literature, as contained in most research papers as part of a background or related work section.
Ad hoc reviews may develop theory \cite[e.g.][]{ralph2018toward}, or integrate a new theory into existing literature \cite[cf.][]{stol2016grounded}. Ad hoc reviews can support a position 
paper, or tertiary scholarship. 

Ad hoc reviews use purposive sampling \cite{baltes2020sampling}; that is, researchers purposefully select papers or studies that are useful, relevant, or support their arguments. Ad hoc reviews are often appropriate, for example, to support theory development, identify promising research topics, or prepare for a comprehensive exam. However, they suffer from important limitations: their unsystematic nature introduces sampling bias and defies replication.
This may lead to cherry-picking of evidence supporting authors' arguments \cite{gough2017introduction}.
Therefore, \textit{ad hoc reviews are inappropriate for supporting empirical statements} such as ``x causes y'', ``most people/objects have property P'', or ``process P has structure S or follows rules R''.

\vspace{-0.5ex}

\subsection{Systematic Reviews}
The term \textit{systematic review} has two meanings, as follows:
\begin{enumerate}
    \item a literature review that employs a systematic (hence the name), replicable process of selecting primary studies for inclusion (below referred to as \textit{systematic review}), including case surveys, critical reviews, meta-analyses,  meta-syntheses, and scoping reviews. 
    \item a systematic review that uses meta-analysis of quantitative studies (especially experiments) to assess the strength of the evidence for specific, usually causal, propositions (below referred to as \textit{meta-analysis}). 
\end{enumerate}

This double meaning is due to the history of systematic reviews. The concept of a meta-analytic systematic review emerged from health and medicine in the late 20\textsuperscript{th} century \cite[ch.2]{purssell2020brief}, when practitioners could not stay current with accelerating research production. When Chalmers~\cite{chalmers1993cochrane} founded the Cochrane Library in 1993,
the medical community coalesced around using meta-analysis to inform evidence-based practice. Now, \textit{systematic review} is often  conflated with \textit{meta-analysis}. However, other kinds of systematic reviews have also been around for decades. Scoping reviews were proposed no later than \citeyear{arksey2005scoping} \cite{arksey2005scoping}. Meta-synthesis goes back at least to \citeyear{jensen1996meta} \cite{jensen1996meta}. Case surveys were proposed as far back as \citeyear{lucas1974case} \cite{lucas1974case}. 

Systematic reviews begin by searching databases for primary studies that meet pre-established criteria. Most types of systematic reviews seek to identify \textit{all} the primary studies that meet the selection criteria by applying various techniques to mitigate sampling bias and publication bias including: ``backward and forward snowballing searches; checking profiles of prolific authors in the area;  searching both formal databases (e.g. ACM Digital Library) and indexes (e.g. Google Scholar); searching for relevant dissertations; searching pre-print servers (e.g. arXiv); soliciting unpublished manuscripts through appropriate listservs or social media; contacting known authors in the area'' \cite{ralph2020acm}.

The way that these primary studies are analyzed determines the \textit{type} of systematic review. 

\subsubsection{Meta-analysis}

An archetypal systematic review analyzes a set of randomized controlled experiments with the same independent and dependent variables.

Suppose, ten different experiments randomized SE undergrads into a control group (who complete tasks individually) and a treatment group (who complete tasks in pairs); that is, individual vs. pair programming. The dependent variable was the number of tasks completed successfully. Each primary study reports the results of an independent samples t-test including the mean and standard deviation for each group, the t-statistic, the p-value, Cohen's $D$ (effect size) and the 95\% confidence interval for $D$. 

Our aim, then, is to combine the results of these ten experiments to estimate the effect size of pair programming on effectiveness. Suppose that four studies found a negative effect, three found no significant effect, and three found a positive effect. We \textbf{DO NOT} proceed by vote counting; that is, conclude that the effect is negative because four negative results outweigh three positive results. Vote counting is invalid because primary studies can have wildly different sizes and quality levels. What if the studies that found positive effects were much larger and more rigorous; while the studies that found negative effects were small and confounded? What if, when the three studies without significant results are aggregated, together their results are significant? 

Instead of vote-counting, we apply meta-analysis \cite{glass1976primary}; that is, we statistically aggregate primary study results into a global effect size estimate. This is often possible with the summary data reported in papers, without the original datasets. 

Meta-analysis can aggregate results from other kinds of (quantitative) methods as long as the studies have the same independent and dependent variables, or overlapping sets of variables. The more complicated the overlaps, the more complicated the meta-analytic model. A comprehensive tutorial on statistical procedures for meta-analysis is beyond the scope of this paper, but readily available~\cite{borenstein2021introduction}.

Meta-analytic reviews essentially have the same research question as the studies being reviewed. The purpose of the meta-analysis is to reach a more reliable, robust conclusion by aggregating all available data, implying two important criteria for meta-analysis: (1) researchers should go to great lengths to find \textbf{ALL} relevant studies; and (2) researchers \textit{must} evaluate the quality of each primary study and either exclude low quality studies or include study quality as a covariate in the meta-analytic model.

In summary, when scientists equate systematic reviews with evidence-based practice, they usually mean meta-analytic reviews. Meta analysis aggregates quantitative studies that investigate the same or overlapping hypotheses. \textit{They do not simply describe} existing research. 
Meta-analysis is rare in software engineering. While there are several good examples \cite[e.g.][]{shepperd2014researcher,hannay2009effectiveness,rafique2012effects}, quality meta-analysis is dwarfed by superficial scoping reviews \cite{cruzes2011research}.

\subsubsection{Meta-synthesis}

Meta-synthesis---aka thematic analysis, narrative synthesis, meta-ethnography, and interpretive synthesis---refers to a family of methods of aggregating qualitative studies \citep{jensen1996meta}. Meta-synthesis is approximately the constructivist analogue of meta-analysis. After identifying the primary studies, the researcher applies hermeneutical and dialectical analyses to understand each primary study, translate them into each other, and construct an account of the body of research; for example, a theory of the central phenomenon that unites the primary studies. 

Meta-synthesis requires expertise in qualitative methods and familiarity with the underlying philosophical assumptions. If one does not know what ``hermeneutical and dialectical analyses'' means, one should not attempt meta-synthesis. Meta-synthesis is \textbf{NOT} organizing papers into categories (as in scoping reviews). Meta-synthesis is synthesizing a credible, nuanced account of a phenomenon from prior qualitative findings.  

In principle, meta-synthesis can be applied to both qualitative and quantitative work. In practice, such combinations are philosophically strained. 

\subsubsection{Case Survey (aka Case Meta-analysis)}

A case survey's primary studies are (typically qualitative) case studies in the broadest sense (i.e., a scholarly account of some events). Experience reports and grey literature may or may not be included, depending on the study's purposes.  
Unlike meta-synthesis, however, a case survey transforms qualitative accounts into a quantitative dataset that supports null-hypothesis testing. Case surveys share the philosophy of meta-analysis (positivism), not meta-synthesis (constructivism). 
They typically begin with a priori hypotheses and an a priori coding scheme. The researcher reads each case and extracts data into the coding scheme, often using simple dichotomous variables like `did the team have retrospective meetings? [yes/no]' or `does the case mention coordination problems? [yes/no]' The resulting dataset is often too sparse for regression modeling, so researchers use simple bivariate correlations to test hypotheses \cite{bullock1987case}. 

Case surveys were proposed by the Rand Corporation as a ``way to aggregate existing research'' \cite{lucas1974case}, quickly picked up by Yin \cite{yin1975using}, and later elaborated in management \cite{bullock1987case,larsson1993case}. Now case meta-analysis is widely used in management and information systems \cite{jurisch2013using}. While SE-specfic case survey guidelines are available~\cite[e.g.][]{melegati2020case,petersen2020guidelines}, SE case surveys remain rare. However, case surveys have been used to investigate strategic pivots at software start-ups \cite{bajwa2017failures} and how organizations select component sourcing options \cite{petersen2017choosing}. 
Case surveys have great potential in SE because case studies are so common. 

\subsubsection{Critical Reviews}

A \textit{critical review} analyzes a sample of primary studies to support an argument or critique.\footnote{The term \textit{critical review} has been used differently elsewhere, but we focus on the meaning in SE.} For example, Stol et al. \citep{stol2016grounded}'s critical review of the use of grounded theory in SE \textit{criticizes} method slurring; that is, claiming to have used a research methodology that was not actually used to create illusory legitimacy. Similarly, Baltes and Ralph's critical review~\citep{baltes2020sampling} \textit{criticizes} how SE researchers often overstate sample representativeness and conflate random sampling with representative sampling. Indeed, critical reviews in software engineering often investigate methodological topics, including how ethnography is reported \citep{zhang2019ethnographic} or how qualitative research is synthesized \citep{huang2018synthesizing}. 

Critical reviews differ from case surveys and meta-analysis in two important ways. First, meta-analytic reviews aggregate evidence regarding causal relationships to generate evidence-based recommendations, while critical reviews critically evaluate issues. Critical reviews are not for supporting evidence-based practice, or summarizing the evidence for a theory. Critical reviews are part of the meta-scientific discourse; i.e., the conversations a scientific community has internally about how it conducts research. 

Second, for many critical reviews, including \textit{all} relevant primary studies is impossible and unnecessary. For example, a critical review of adherence to the \textit{Introduction, Method, Results and Discussion framework} (IMRaD) framework~\cite{sollaci2004introduction} could include all SE papers ever written. Instead, a random sample of papers from a selection of top journals and conferences is sufficient because critical reviews do not assess causal claims so publication bias---``what if significant results were published but non-significant results were not?''---is irrelevant. 

Analysis performed within a critical review can be quantitative, qualitative or both. However, critical reviews typically adopt a \textit{critical} stance; that is, they go beyond mere description and offer specific critiques of the work being reviewed. 

\subsubsection{Scoping Reviews (aka Systematic Mapping Studies)}

What is often called a ``systematic mapping study'' in SE~\cite{petersen2008systematic} is usually called a \textit{scoping review} elsewhere (e.g. in health, medicine, and psychology). 
The purpose of a scoping review is to understand the status of research on a particular topic, typically by mapping primary studies into categories (hence, ``mapping study''). 

Scoping reviews are often primarily descriptive. They count the number of studies on a topic. They often organize studies by research method, subtopic, authors, geographical location, publication venue, etc. They often conclude that more research is needed in this or that subtopic. For example, Mohanani et al. \cite{mohanani2018cognitive} mapped primary studies according to which cognitive bias (e.g. confirmation bias) they investigated and in which area of software development (e.g. design, management) they investigated it, then called for more research on \textit{debiasing} (preventing or mitigating cognitive biases). 

The problem with scoping reviews is that they typically fail to fulfill any of the core purposes of systematic reviews:

\emph{Meta-analysis} and \emph{case surveys} synthesize the results of many studies to answer specific empirical (often causal) questions about the world. Scoping reviews include a similar search but typically do not provide sufficient quantitative synthesis to answer important empirical questions. Therefore, scoping reviews do not inform evidence-based practice like meta-analyses and case surveys do. 

\emph{Meta-synthesis} involves deep, theory-oriented reinterpretation of related qualitative studies. While scoping reviews often include qualitative analysis (e.g. mapping or categorization), that analysis is often too superficial to generate novel, useful theories.

\emph{Critical reviews} use a sample of papers to demonstrate an important pattern for a scientific community's internal discourse. 
While scoping reviews often give recommendations regarding future research, they focus on an empirical topic (e.g. cognitive biases in SE); not a meta-scientific topic (e.g. construct validity); therefore, they are not configured, from the outset, to deliver useful meta-scientific critique.

In summary, scoping reviews begin like other kinds of systematic reviews, but stop short of \textit{synthesizing} the data into aggregate empirical results, theory, or meta-scientific critique. This is by definition: if a scoping review applies a meta-analytic model to aggregate primary study results, or applies hermeneutics and dialectics to synthesize qualitative accounts, or develops an evidence-based critique of a scientific practice, it is no longer a scoping review; it is a meta-analysis, a case survey, a meta-synthesis, or a critical review. Some authors therefore recommend a scoping review ``as a precursor to a systematic review'' \cite{munn2018systematic}. 

\subsubsection{Rapid Reviews}

A \textit{rapid review} is a meta-analysis that makes methodological compromises to reduce completion time~\cite{ganann2010expediting}. Ganann et al. found many such compromises including restricting the literature search, truncating results, omitting techniques for overcoming publication bias (e.g. reference snowballing); streamlining screening and data extraction, and skipping quality assessment \cite{ganann2010expediting}. 

Rapid reviews are justified \textit{if and only if} evidence is needed to support imminent decisions, and waiting for a comprehensive meta-analytic review would be harmful. These conditions occur in health and medicine, for example, when an unprecedented viral pandemic strikes. These conditions do not occur frequently in SE. The term \textit{rapid review} should not be used to legitimize bad systematic reviews where there is no pressing need for immediate results. Moreover, SE-related topics rarely have so many primary studies that it would take more than a year to complete a comprehensive review.





\section{Discussion}

Secondary research that synthesizes the results of a body of work into robust, reliable findings and practical ramifications is crucial for evidence-based practice. Secondary research is the funnel that concentrates the overwhelming flow of empirical results into something practitioners can digest and use. 
However, most secondary research in SE contains no synthesis of findings \cite{cruzes2011research}, because most SE secondary studies are scoping reviews. Scoping reviews struggle to get from a broad description of what is happening in a field to specific recommendations for practitioners. While the purpose of meta-analysis is to answer a specific empirical question by combining the results of all of the studies that tested corresponding hypotheses, scoping reviews just classify studies. 

Meanwhile, SE researchers perceive inadequate numbers and quality of primary studies as major barriers to good secondary research \citep{guzman2014survey}. Useful meta-analysis (and meta-synthesis) are only possible given sufficiently large bodies of similar quantitative (and qualitative) studies. With their sparse data sets, case surveys need \textit{many} case studies.

\subsection{Recommendations}
\label{section:recommendations}
Our primary recommendation is as follows:

\smallskip
{\narrower \noindent \textit{\textbf{Recommendation:} Software engineering needs \textit{more} meta-analysis, case surveys, and meta-synthesis, but \textit{fewer} scoping and rapid reviews.} \par}
\smallskip

If the body of evidence is sufficient for a meta-analysis, case survey, or meta-synthesis, by all means, do one. Top venues should welcome competent meta-analyses, case surveys, and meta-syntheses as these approaches are central to evidence-based practice. In contrast, predominately descriptive scoping reviews are essentially works-in-progress. They belong in workshops, poster sessions and short paper tracks. If the literature does not support a meta-analysis, case survey, or meta-synthesis, then a scoping review is unlikely to substantively contribute to the scholarly discourse. If the literature does support some kind of synthesis, the scoping review is like a pilot study for a more ambitious systematic review.

Rapid reviews similarly have little place in SE. The conditions that justify a rapid review rarely occur in our field. Top venues should expect rapid reviews to justify methodological compromises and explain why results are needed immediately. The term ``rapid review'' should not be used to legitimize a bad meta-analysis.  

The case for critical reviews is more complicated. SE does not need more critical reviews per se; SE needs a more vivid, robust meta-scientific discourse. Meta-science in SE has been hamstrung by reviewers and venues insisting that every paper reports an empirical study, regardless of whether its research question is empirical. As long as this unwise position remains, we need critical reviews to inject meta-science into our publications. 

Since published guidelines for critical reviews are scarce, we will state some recommendations. First, when elucidating a ubiquitous problem (e.g. poor construct validity) citing specific examples of papers making that mistake is unnecessary. The credibility gained is not worth the costs to community cohesion, interpersonal relationships, and future collaborations. Authors should explain this reasoning and reviewers should accept this compromise. Second, reviewers should not misapply the ``find ALL relevant studies'' or ``exclude low quality studies'' criteria to critical reviews. 

Finally, authors should identify the kind of systematic review they have done. Avoid using ``systematic review'' as a synonym for a meta-analysis or a scoping review. In the broadest sense, \textbf{any review that uses a systematic process of identifying primary studies is a systematic review}. 

\subsection{Limitations}
This is not an empirical paper. It is meta-science: part of our community's internal discourse about what we are doing, why, and how to do it. This paper is credible because it meets the expectations laid out in the Empirical Standard for Meta-science \cite{ralph2020acm}: it is useful because most SE researchers read or perform literature reviews; it makes clear recommendations (Section \ref{section:recommendations}); presents coherent arguments; ``goes beyond summarizing methodological guidance from existing works... [and]  provides insight specifically for software engineering''. It is limited by excluding tertiary studies and providing relatively brief descriptions of each type of review to respect the venue's page limit. Some reviewers of this paper argued that it did not fit the call-for-papers. This criticism underlines the tendency in SE to abandon a meta-scientific discourse. 
We envision a future where a lively meta-scientific discourse is an essential part of our research community.

\section{Conclusion}

\textit{Systematic review} is often conflated with \textit{meta-analysis}. We recommend using \textit{systematic review} as the category of secondary research in which primary studies are selected according to a systematic, replicable process; that is, the opposite of an \textit{ad hoc} review. Meta-analysis, case survey, meta-synthesis, critical review, scoping review, and rapid review are all types of systematic review. 

SE needs more meta-analysis, meta-synthesis, and case surveys. These methods synthesize bodies of related work into actionable recommendations for practitioners and are therefore crucial for evidence-based practice. Scoping reviews, in contrast, rarely include enough synthesis to support evidence-based practice. We consider scoping reviews a prelude to a more ambitious project. If published at all, scoping reviews belong with other works-in-progress in posters, workshops or short papers. Rapid reviews, meanwhile, are a valuable tool but SE phenomena rarely manifest the conditions under which rapid reviews are justified. Finally, critical reviews have different best practices and reviewers should stop misapplying meta-analysis criteria to critical reviews.    

We hope this paper helps researchers distinguish between different types of literature reviews---paving the way for more mature secondary research in software engineering.



\balance
\bibliographystyle{ACM-Reference-Format}
\bibliography{bib}



\end{document}